# The role of pairing correlations for the description of $\beta$ vibrations at superdeformed shapes[*]

Piotr Magierski[†] and Ramon Wyss

Physics Department Frescati, Royal Institute of Technology, S-10405 Stockholm, Sweden

In this paper we analyze the possibility of the appearance of quadrupole vibrational states in superdeformed nuclei from the $A \approx 150$ region. We discuss the role of pairing correlations for a proper description of quadrupole vibrations, especially of the $\beta$ type.

The theoretical study of $\beta$-vibrations of deformed and rotating nuclei have so far been addressed rather rarely. The $\beta$ vibrations are known to couple strongly with the pairing modes of the nucleus [1, 2]. Hence, to obtain their correct description requires the proper treatment of pairing correlations in nuclei. This fact is especially important for the high spin states where the standard BCS theory breaks down. Another problem is associated with the fact that the pairing interaction in the form it is used in most applications (i.e. monopole pairing) violates the Galilean invariance. This implies that the dependence of the $\beta$-vibrational states on the pairing interaction contains an unphysical component.

In the present paper we would like to analyze the influence of pairing correlations on the properties of $\beta$-vibrational states in both normally deformed as well as superdeformed nuclei from the mass region $A \approx 150$. This investigation was inspired by recent reports indicating the existence of a $\beta$-vibrational band in superdefomed $^{150}Gd$ [3] and $^{148}Gd$ [4].

The study has been performed in the QRPA approximation using the Woods-Saxon single particle states as a basis. The inclusion of the pairing correlations in the mean-field has been achieved by solving the HFB equations. The details of these calculations were described elsewhere [5]. The

---

[*] Presented at the XXXIII Zakopane School of Physics
[†] On leave of absence from Institute of Physics, Warsaw University of Technology, Warsaw, Poland.

QRPA method aims at finding the excited states of the Hamiltonian:

$$\hat{H} = \hat{H}_{qp} + \hat{H}_{int} = \sum_{\mu} E^{\omega}_{\mu} \alpha^+_{\mu} \alpha_{\mu} + \frac{1}{2} \sum_i \chi_i \hat{Q}^+_i \hat{Q}_i, \qquad (1)$$

where $\alpha$, $\alpha^+$ are the quasiparticle operators, and $E^{\omega}$ are quasiparticle energies. The superscript $\omega$ is used to describe the rotating system. The perturbation of the single quasiparticle Hamiltonian is caused by a separable interaction where $\hat{Q}_i$ denote the following operators: (i) quadrupole operators in the particle-hole channel: $\hat{Q}_{2\mu}$, (ii) monopole pairing operator: $\hat{P}_{00}$, (iii) quadrupole pairing operators: $\hat{P}_{2\mu}$.

Since we study deformed systems we use the double-stretched operators which represent a remarkable improvement over the conventional modes in the sense that they satisfy the self-consistency between the nucleonic density and the potential. Another advantage is that the selfconsistent strengths of the multipole interactions do not depend on the $K$ quantum number [6].

In our calculations we have adjusted the strength of the interaction in the particle-hole channel to obtain the spurious $1^+$ mode at zero energy. It should be emphasized that the determination of the coupling constant at zero frequency is in general not valid for high spin states. We can assume that such an approximation is justified if the deformation along the band does not differ considerably. That is in fact the case for many well developed rotational structures in nuclei, in particular for the superdeformed bands. For all nuclei we discuss in this paper, the change of the energy of the spurious mode along the rotational band does not exceed 150 keV.

The value of the quadrupole pairing interaction was adjusted to restore the Galilean invariance in the RPA approximation [7, 8].

The RPA equations can be expressed in the form:

$$[\hat{H}_{qp} + \hat{H}_{int}, \hat{X}_n]_{RPA} = \hbar \Omega_n \hat{X}_n, \qquad (2)$$

where $\hbar\Omega_n$ are the excitation energies. The operator $\hat{X}_n$ has the form:

$$\hat{X}_n = \sum_{\mu<\nu} (\phi^n_{\mu\nu} \alpha^+_{\mu} \alpha^+_{\nu} - \psi^n_{\mu\nu} \alpha_{\nu} \alpha_{\mu}), \qquad (3)$$

where $\phi^n$ and $\psi^n$ represent forward and backward amplitudes of the solution, respectively. Equation (2) for separable forces is equivalent to:

$$det\left(\frac{1}{\chi_i}\delta_{ij} + R_{ij}(\Omega)\right) = 0, \qquad (4)$$

with a multidimensional response function $R$.

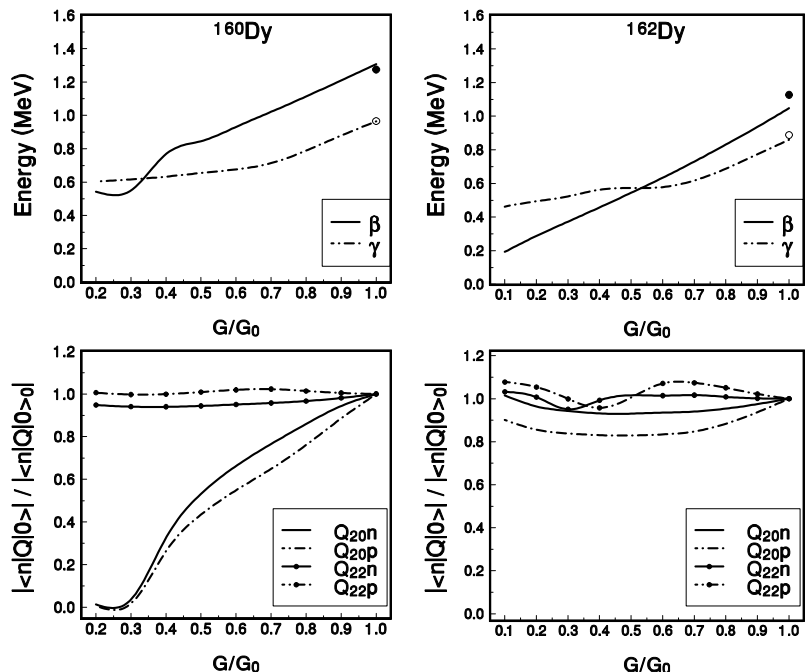

Fig. 1. The excitation energy of the $\beta$ and $\gamma$ vibrations as a function of the pairing strength $G/G_0$ for two nuclei: $^{160}Dy$ and $^{162}Dy$. Open and filled circles denote the experimental $\gamma$ and $\beta$ vibrational states, respectively. In the lower part the relative transitions matrix elements for protons and neutrons are shown (see legend).

In order to illustrate the dependence of the $\beta$ vibration on pairing correlations in the mean-field we consider first normally deformed nuclei (see Fig.1). The transition matrix elements were normalized to their values at normal pairing strength $G_0$. In this calculation the coupling of the $\beta$ and $\gamma$ modes to the pairing modes has not been taken into account. It is clearly visible that in both cases the $\beta$-vibrational states are more sensitive to pairing correlations in the mean field. Note the nice agreement between experiment and theory for the vibrational states.

We consider now superdeformed state of $^{150}Gd$ at the rotational frequency $\hbar\omega = 0.55$ MeV. In this case the normal HFB theory breaks down giving only trivial unpaired solutions. In such a case no collective $\beta$ vibrations can be obtained. In order to take into account the effect of particle number fluctuations we employ the Lipkin-Nogami method. Although the method is not consistent with the QRPA approach it can be treated as a tool for generating a finite pairing gap. In figure 2 we have shown the behavior of the excitation energy and transition matrix elements as a function of the monopole pairing strength. One can notice that the collectivity of the band is increasing with inreasing pairing strength. For weak pairing correla-tions in the mean field the $\beta$-vibrational state looses its collective character. At $G = 0$ the calculations show only slightly perturbed two-quasiparticle excitations present. It also turns out that the coupling to pairing modes does not influence the result significantly. In figure 3 we have shown the $\beta$-vibrational bands obtained for the case of: (i) no coupling to the pairing modes (dotted line), (ii) coupling to the monopole pairing (dashed line), (iii) coupling to both monopole and quadrupole pairing modes (bold solid line). The results shows that the coupling with the pairing modes produces only a slight shift of the excitation energy of the $\beta$-vibrational state. The character of the state determined by the transition matrix elements remains unchanged.

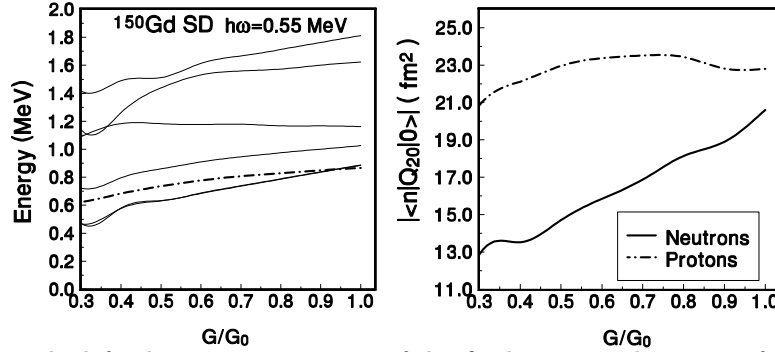

Fig. 2. To the left, the excitation energy of the $\beta$ vibration is shown as a function of the pairing strength (dashed-dotted line). Thin solid lines denote unperturbed two-quasiparticle excitation energies. To the right, transition matrix elements have been plotted in a similar manner. Only the results for the favoured signature quantum number have been shown.

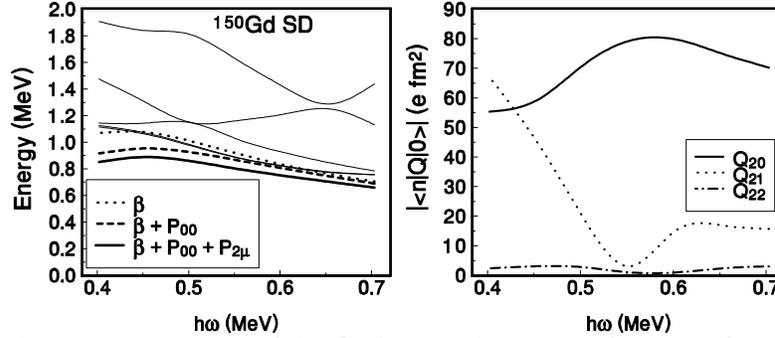

Fig. 3. The excitation energy of the $\beta$ vibrational state as a function of rotational frequency (left figure). The right figure shows the transition matrix elements (for further informations see text). Only the results for the favoured signature quantum number have been shown.

A similar result is obtained for the SD bands in $^{148}Gd$ and $^{150}Dy$.

The following conclusions can be drawn from the presented calculations:

- There is an indication for the presence of collective $\beta$ vibrational states built on the superdeformed states in the $A \approx 150$ mass region.

- The presence of pairing correlations in the mean field is a necessary condition for the existence of $\beta$ vibrational states in $A \approx 150$ region. Hence, the observation of $\beta$-bands at SD shapes will confirm the importance of pairing correlations at high angular momenta.

- In order to obtain a proper description of $\beta$ vibrational bands for superdeformed states the method is needed to take into account pairing correlations beyond the BCS limit. A possible recipe compatible with the QRPA method is to perform selfconsistent RPA calculations using the pairing interaction only, and then construct the RPA method with a particle-hole interaction.

- The coupling of the pairing modes to the $\beta$ vibrations does not produce a large correction to neither the excitation energy nor transition matrix elements at SD shapes.

Helpful discussions with Dr. Hideo Sakamoto are acknowledged. PM acknowledge the financial support from the Swedish Institute and the Göran Gustafsson fundation.